\documentclass[twocolumn,showpacs,preprintnumbers,amsmath,amssymb]{revtex4}

\usepackage{graphicx}
\usepackage{dcolumn}
\usepackage{bm}
\usepackage{color}

\begin{document}


\title{Detection of an Unconventional Superconducting Phase in the Vicinity of the Strong First-Order Magnetic Transition in CrAs Using $^{75}$As-Nuclear Quadrupole Resonance}

\author{Hisashi Kotegawa, Shingo Nakahara, Rui Akamatsu, Hideki Tou, Hitoshi Sugawara, and Hisatomo Harima}

\affiliation{
$^1$Department of Physics, Kobe University, Kobe 657-8501, Japan 
}

\date{\today}

\begin{abstract}

Pressure-induced superconductivity was recently discovered in the binary helimagnet CrAs. We report the results of measurements of nuclear quadrupole resonance for CrAs under pressure. In the vicinity of the critical pressure $P_c$ between the helimagnetic (HM) and paramagnetic (PM) phases, a phase separation is observed. The large internal field remaining in the phase-separated HM state indicates that the HM phase disappears through a strong first-order transition. This indicates the absence of a quantum critical point in CrAs; however, the nuclear spin-lattice relaxation rate $1/T_1$ reveals that substantial magnetic fluctuations are present in the PM state. The absence of a coherence effect in $1/T_1$ in the superconducting state provides evidence that CrAs is the first Cr-based unconventional superconductor.

\end{abstract}

\maketitle

In a conventional Bardeen-Cooper-Schrieffer (BCS) superconductor, electrons form Cooper pairs through negative interactions via an electron-phonon coupling.
This gives a sign-uniform $s$-wave character to the superconducting (SC) gap symmetry.
If positive interactions work electron pairs to induce superconductivity, unconventional superconductivity with a sign-changing order parameter is realized beyond the well-established simple $s$ wave.
This fascinating form of superconductivity has been discovered in specific materials, such as heavy fermion systems, high-$T_c$ cuprates, organic systems, ruthenate, and recent Fe-based superconductors.
A common feature of these materials is that superconductivity appears due to the instability of some degree of freedom, which is mainly induced on the verge of a magnetically ordered phase.

Very recently, Wu {\it et al.} and Kotegawa {\it et al.} independently discovered pressure-induced superconductivity in CrAs in the vicinity of the helimagnetic (HM) phase \cite{Wu2,Kotegawa}.
This is the first example of superconductivity found in a Cr-based magnetic system.
CrAs has a MnP-type orthorhombic crystal structure with a $Pnma$ space group.
This structure possesses a space-inversion symmetry, but it is locally missing at the Cr and As sites, and both of these atoms form a zigzag chain along the $a$-axis.
The magnetic transition of the first order into a double-helical state occurs at $T_N \sim 265$ K.
This transition is accompanied by a large magnetostriction, and a lowering of the crystal structure symmetry has not been reported \cite{Boller,Suzuki}.
The propagation vector is incommensurate with $0.354 \cdot 2\pi c^*$, where $c^*$ is a unit vector along the $c$ axis, and magnetic moments of $\sim1.7\mu_B$/Cr lie in the $ab$ plane \cite{Watanabe,Selte}.
The application of pressure drastically suppresses $T_N$, and the HM phase disappears above a critical pressure of $P_c\sim0.7$ GPa \cite{Wu2,Kotegawa,Zavadskil}.
Superconductivity appears together with the suppression of the HM phase, showing a maximum SC transition temperature of $T_c\sim2.2$ K at $\sim1.0$ GPa, after which $T_c$ decreases gradually with increasing pressure.
The pressure-temperature phase diagram of CrAs with a first-order magnetic transition is reminiscent of those of some pressure-induced Fe-based superconductors, such as SrFe$_2$As$_2$ \cite{Kotegawa_SrFe2As2,Kitagawa}.
Magnetic interactions are conjectured to play a crucial role in the induction of superconductivity in CrAs.
Experimental investigation is desired to elucidate how SC symmetry is realized in CrAs.

Here, we report the results of nuclear quadrupole resonance (NQR) measurements under ambient and high pressures to elucidate the underling electronic correlations and SC symmetry in CrAs.
From the nuclear spin relaxation rate $1/T_1$, we obtained a strong suggestion that the superconductivity realized under moderate magnetic fluctuations of CrAs is of an unconventional nature.

Single crystals of CrAs were prepared by the Sn-flux method as described in Ref. \cite{Kotegawa}.
The NQR measurements under pressure were performed by using a piston-cylinder cell and Daphne 7474 as the pressure-transmitting medium \cite{Murata}.
The applied pressure was estimated from the $T_{c}$ of the lead manometer.
Several as-grown single crystals were put into the pressure cell, and NQR measurements were performed by using the conventional spin-echo method for the $^{75}$As nucleus.
A $^3$He cryostat was utilized to perform NQR measurement at low temperatures.
To obtain $T_1$ in the paramagnetic (PM) state, the recovery curves obtained at the ($\pm1/2 \leftrightarrow \pm3/2$) transition are fitted by a single exponential function.

\begin{figure}[htb]
\centering
\includegraphics[width=0.8\linewidth]{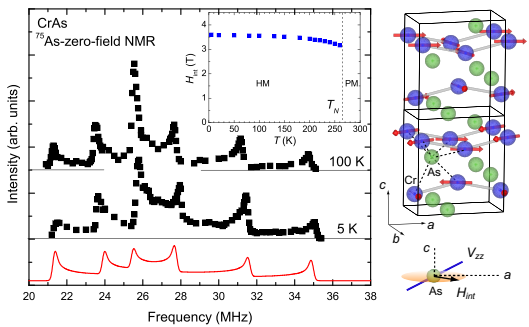}
\caption[]{(color online) Zero-field NMR spectra for the $^{75}$As nucleus at ambient pressure. The complicated spectrum at 5 K can be almost reproduced by a model based on the HM state reported by previous neutron scattering measurements, as shown by the red curve. The crystal and magnetic structures are shown. The As site is surrounded by six neighboring Cr sites composed of three inequivalent sites for the As site. The inset shows the temperature dependence of $H_{int}$ at the As site, which decreased gradually with increasing temperature and remained large at $T_N$ due to the first-order transition.
}
\end{figure}

Figure 1 shows the $^{75}$As NQR- or the so-called zero-field nuclear magnetic resonance (NMR)-spectra in the HM state at ambient pressure, where the energy levels of the nuclear spins are mainly lifted by the internal field.
Because the nuclear spin of $^{75}$As is $I=3/2$, and because all the As sites in the crystal are equivalent to each other, three transitions were expected for a commensurate magnetic structure, whereas complicated spectra composed of six peaks were obtained at both 5 and 100 K.
Each peak being not well separated suggests the presence of a spatial modulation of the internal field at the As sites.
In fact, previous neutron scattering measurements have shown that the magnetic structure is incommensurate \cite{Watanabe,Selte}.
Because the measurements also showed that the magnetic moments lay in the $ab$ plane, we performed a simulation based on this magnetic structure; that is, we assumed that the internal field $H_{int}$ at the As site lies in the $ab$ plane through dominant isotropic hyperfine coupling between the Cr moments and the As nuclei. Because of the incommensurate structure of the Cr moments, the direction of the $H_{int}$ rotates in the respective As sites.
The EFG parameter at the As site in the HM state is unknown; however, crystallographically, one of the principal axes is restricted to be along the $b$ axis.
Through trial and error, we obtained the red curve to reproduce the experimental result, where the distribution of $\nu_Q$ was considered by a convolution of a Lorentzian whose width $\Delta \nu_Q=0.1$ MHz.
The result gave the quadrupole frequency as $\nu_Q=14.7$ MHz, the asymmetry parameter as $\eta=0.99$, $H_{int}^a=3.82$ T along the $a$-axis, and $H_{int}^b=3.35$ T along the $b$ axis.
The direction of the first principal axis of the EFG, $V_{zz}$, was tilted from $a$ axis towards the $c$ axis at 29.8$^{\circ}$, and the third principal axis lay along the $b$ axis.
The anisotropy of $H_{int}$ in the $ab$ plane is considered to originate in the small anisotropy of the hyperfine coupling constant.
The inset shows the temperature dependence of $H_{int}=(H_{int}^a+H_{int}^b)/2$, which is estimated from the peak frequency $f_{peak}$ at $\sim26$ MHz through $H_{int}=(f_{peak}+0.41)/\gamma_N$, where $\gamma_N=7.292$ MHz/T is a gyromagnetic ratio for the As nucleus.
The shift of 0.41 MHz is a contribution from the quadrupole interaction, which is small compared with $H_{int}$ and assumed to be independent of temperature.
The $H_{int}$ decreases gradually with increasing temperature and vanishes suddenly at $T_N$ because of a first-order phase transition.

\begin{figure}[htb]
\centering
\includegraphics[width=0.6\linewidth]{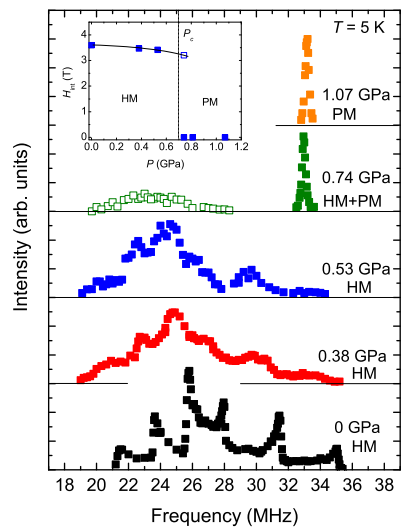}
\caption[]{(color online) Pressure dependence of spectra at 5 K. The spectrum is broadened by applying pressure, but its structure is similar up to 0.53 GPa. At 0.74 GPa close to $P_c$, the signal from the HM phase is suppressed and the new signal corresponding to the PM phase appears, indicating that the phase separation between the HM and PM phases occurs in the vicinity of $P_c$. The intensity of the HM phase at 0.74 GPa is $\sim5$ times smaller than that at 0.53 GPa. The inset shows the pressure dependence of $H_{int}$, which has a large value on the border of the HM-PM transition as well as the case of the temperature dependence shown in the inset in Fig.~1. The dotted line represents $P_c$, as determined by resistivity measurement \cite{Kotegawa}.
}
\end{figure}

Figure 2 shows the pressure dependence of zero-field NMR and NQR spectra measured at 5 K.
The spectrum is remarkably broadened under pressure, probably owing to a distribution of the EFG, but the spectral shape remains similar.
There is no clear signature of a change in the magnetic structure under pressure.
The overall spectrum moves to the low-frequency side with increasing pressure due to a reduction of $H_{int}$.
At 0.74 GPa, the intensity of the HM state is drastically suppressed, whereas another sharp spectrum corresponding to the PM state appears at $\sim33$ MHz, which is the ($\pm1/2 \leftrightarrow \pm3/2$) transition.
Since $\eta$ in the PM state is unknown, the $\nu_Q$ is estimated to be in the range between $28.6$ (for $\eta=1$) and $33$ MHz (for $\eta=0$).
The $\nu_Q$ in the PM state is about twice larger than that in the HM state, owing to the large change in the crystal structure and the possible change in the Fermi surface \cite{Boller,Suzuki}.
We observed simultaneously the spectra for both the HM and PM states at 0.74 GPa, giving microscopic evidence of a phase separation.
We did not observe the signal from the HM state at 0.81 GPa; thus, the homogeneous PM phase was realized above approximately 0.8 GPa.
The inset shows the pressure dependence of $H_{int}$, which is estimated from the peak frequency by the same procedure as that in the inset in Fig.~1 by assuming the small quadrupole contribution is unchanged under pressure.
The $H_{int}$ gradually decreases under pressure and maintains a large value even in the vicinity of $P_c$.
We cannot estimate the accurate ordered moment under pressure, because the pressure dependence of the hyperfine coupling constant is uncertain; however, the large $H_{int}$ in the vicinity of $P_c$ indicates that the HM-PM transition at low temperatures is of strong first order.
Our observations suggest that CrAs does not possess a quantum critical point (QCP), which is a continuous phase transition at 0 K, on the pressure-temperature phase diagram.

\begin{figure}[htb]
\centering
\includegraphics[width=0.8\linewidth]{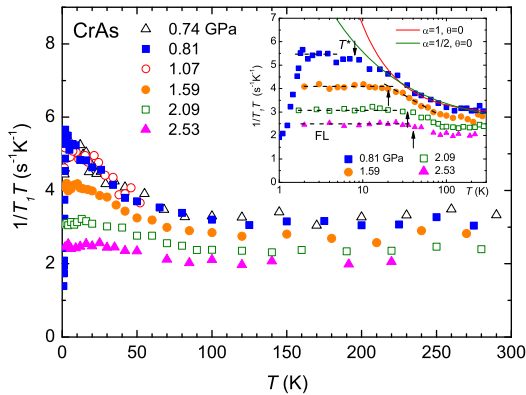}
\caption[]{(color online) Temperature dependence of $1/T_1T$ in the PM state, where superconductivity occurs. The increase in $1/T_1T$ toward low temperatures demonstrates the development of magnetic fluctuations. The inset shows the same data between 0.81 and 2.53 GPa in a logarithmic scale. Even at 0.81 GPa, $1/T_1T$ deviates from the behavior expected at a QCP  below $\sim20$ K. The characteristic temperature $T^*$ was determined from an intersection point of each extrapolation from the low-temperature and high-temperature sides, as shown for the data at 1.59 GPa.
}
\end{figure}

Figure 3 shows the temperature dependence of $1/T_1T$ in the PM phase under pressure.
$1/T_1T$ is almost constant for temperatures from $300$ to $100$ K and increases toward lower temperatures below $\sim100$ K.
The deviation from Fermi liquid (FL) behavior of $T_1T=const$ below $\sim100$ K clearly demonstrates the presence of magnetic correlations developing toward lower temperatures.
The enhancement of $1/T_1T$ toward low temperatures has been observed in other $d$-electron superconductors adjoining magnetic phases such as cobalt oxyhydrate \cite{Fujimoto,Ihara,Kusano} and many Fe-based superconductors \cite{Fukazawa,Imai,Nakai_Ba,Oka,Kitagawa}.
In such itinerant magnetic systems, $1/T_1T$ is generally composed of two contributions as follows:
\begin{equation}
\frac{1}{T_1T} = \frac{A}{(T+\theta)^{\alpha}} +  B
\end{equation}
Here, $A$ and $B$ are temperature-independent constants.
The first term is given by spin correlations developing at a specific wave vector.
In a self-consistent renormalization theory \cite{Moriya}, the 3D ferromagnetic and 2D antiferromagnetic correlations give $\alpha=1$, and the 3D antiferromagnetic case gives $\alpha=1/2$.
$\theta=0$ means that the system is located at a QCP.
The second term is obtained from a uniform susceptibility and from a relaxation through orbital angular momentum, which are both related to the density of states at the Fermi level and are independent of temperature in typical metals.
In cobalt oxyhydrate \cite{Ihara,Kusano} and some Fe-based superconductors \cite{Fukazawa,Nakai_Ba,Oka}, low $\theta$s close to zero are obtained, suggesting that these systems are located in the vicinity of the quantum criticality.
Unlike these systems, however, the above form cannot reproduce $1/T_1T$ for CrAs in a wide temperature range down to $T_c$ because the increase of $1/T_1T$ is suppressed at low temperatures.
Consequently, the low $\theta$ cannot be obtained in CrAs from fitting for the data at low temperatures.
The curves shown in the inset in a logarithmic scale were obtained from the fitting the data at 0.81 GPa in a high-temperature range of $20-280$ K with fixing $\theta=0$ \cite{theta}.
We cannot determine $\alpha$ from this fitting, but both curves reproduce the data well above $\sim20$ K.
If the system is close to a QCP, $1/T_1T$ should obey these curves, but the deviations are clear at low temperatures.
The temperature dependence of $1/T_1T$ below $\sim10$ K is weak and close to a constant at 0.81 GPa.
The clear FL behavior above 1.59 GPa suggests that the suppression of $1/T_1T$ corresponds to a crossover behavior from a high-temperature Curie-Weiss or non-Fermi liquid (NFL) regime to a low-temperature FL regime.
We determined the characteristic temperature of the crossover $T^*$ from an intersection point of each extrapolation from the low-temperature and high-temperature sides.
Figure 4 shows the pressure-temperature phase diagram of CrAs, where $T_N$, $T_c$, and $T^*$ are plotted.
The $T^*$ increases with increasing pressure, and the FL state is stable under higher pressures.
A finite $T^*$ in the vicinity of $P_c$ suggests the absence of a QCP in CrAs.
Superconductivity is gradually suppressed with increasing pressure, accompanied by the suppression of the magnetic fluctuations.

\begin{figure}[htb]
\centering
\includegraphics[width=0.8\linewidth]{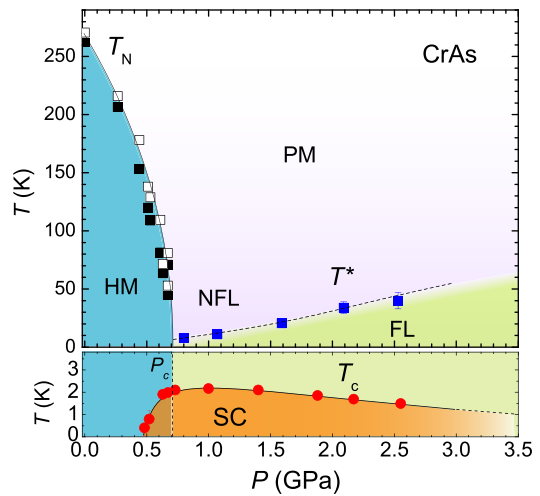}
\caption[]{(color online) Pressure-temperature phase diagram of CrAs. The $T_N$ and $T_c$ are taken from Ref.\cite{Kotegawa} and Supplemental Material. The closed (open) squares represent $T_N$ obtained during the cooling (warming) process. These observations of hysteresis and the present NQR spectrum evidence that the HM-PM transition is of the first order in all pressure-temperature ranges. Even in the vicinity of $P_c$, the system is close to the FL at low temperatures.
}
\end{figure}

\begin{figure}[htb]
\centering
\includegraphics[width=0.8\linewidth]{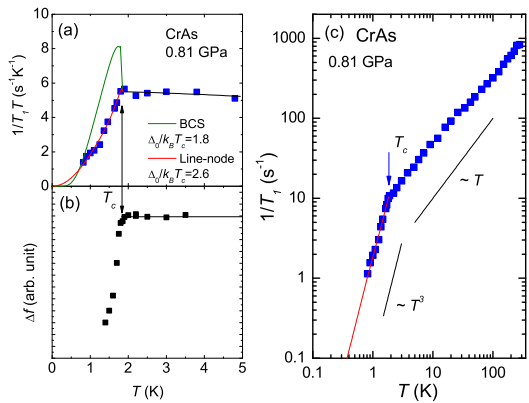}
\caption[]{(color online) (a) Temperature dependence of $1/T_1T$ below 5 K measured under a zero field. The clear reduction in $1/T_1T$ is observed below $T_c=1.85$ K, which is determined by the change in the resonance frequency shown in Fig.~4 (b). The green curve indicates a calculation based on a conventional BCS, whereas the red curve indicates one based on a line-node model (polar type). $1/T_1T$ is clearly reproduced by the unconventional model. (c) Temperature dependence of $1/T_1$ at 0.81 GPa. The red curve based on the line-node model is also plotted. The temperature dependence of $1/T_1$ is close to $T^{3}$ below $T_c$.
}
\end{figure}

Now, we discuss $1/T_1$ in the SC state. 
Figure 5(a) shows the temperature dependence of $1/T_1T$ below 5 K measured at 0.81 GPa.
The temperature dependence of the change in the resonance frequency measured by an $in$-$situ$ NMR coil is also shown in Fig.~5(b).
It clearly shows a diamagnetic signal below $T_c=1.85$ K.
Below $T_c$, $1/T_1T$ shows a clear reduction without the signature of a coherence peak (Hebel-Slichter peak), which is a marker of a BCS superconductor \cite{Hebel}.
The green curve in the figure indicates calculated $1/T_1T$ for a conventional BCS model, which has an isotropic gap of $\Delta_0/k_BT_c = 1.8$.
Evidently this does not match the experimental result below $T_c$.
On the other hand, the red curve indicates calculated $1/T_1T$ for a line-node model [polar type; $\Delta(\theta,\phi)=\Delta_0 \cos \theta$] with $\Delta_0/k_BT_c = 2.6$ and reproduce the data well, meaning that the coherence effect is completely absent in CrAs and suggesting strongly that superconductivity is not of the conventional BCS but unconventional in nature.
CrAs is the first example of unconventional superconductivity seen in Cr-based systems.


Here, we used a polar-type line-node model as a simple example, which gives $T^3$ dependence in $1/T_1$ at low temperatures, but the SC gap function is not restricted by the present experiment.
For example, sign-changing nodeless $s^{\pm}$-wave symmetry is considered to be realized in some of the Fe-based multigap superconductors, and this symmetry can reproduce the $T^3$ dependence of $1/T_1$ \cite{Parker,Chubukov,Nagai}.
In fact, our band calculation , shown in Supplemental Material, suggests that the Fermi surface of CrAs is composed of several sheets.
The present NQR results strongly suggest that a sign in the SC gap must change somewhere in the momentum space; thus, the possibility of a nodeless gap cannot be excluded.

In summary, we performed NQR measurements on pressure-induced superconductor CrAs.
The phase separation between the HM and PM phases in the vicinity of $P_c$ and the insensitivity of $H_{int}$ against pressure provide microscopic evidence of strong first-order HM-PM transition at $P_c$.
Although this indicates that CrAs does not possess a QCP, $1/T_1T$ in the PM state increases significantly from 100 to $\sim20$ K, suggesting that substantial magnetic fluctuations remain in the PM phase.
Below $\sim20$ K, $1/T_1T$ approaches the FL behavior, and superconductivity sets in.
$1/T_1T$ does not show a coherence peak, and decreases sharply below $T_c$, giving a clear indication that superconductivity in CrAs is of an unconventional nature.
This first example of unconventional superconductivity in a Cr-based system opens a new route to the development of the field of superconductivity.

\section*{Acknowledgments}

We thank Haruki Matsuno for the experimental support.
This work has been supported in part by Grants-in-Aid for Scientific Research (Nos.  22740231, 20102005, and 24340085) from the Ministry of Education, Culture, Sports, Science and Technology (MEXT) of Japan.

\clearpage
\setcounter{figure}{0}

\begin{center}

{\bf Detection of an Unconventional Superconducting Phase in the Vicinity of the Strong First-Order Magnetic Transition in CrAs Using $^{75}$As-Nuclear Quadrupole Resonance \\
  - Supplemental Material-}

\end{center}

\subsection{Pressure dependence of the superconducting transition temperature}

Figure 1 shows the temperature dependence of the resistivity for CrAs under pressure.
The data above 1.00 GPa have been already published in Ref. [2].
At 0.48 and 0.52 GPa in the helimagnetic (HM) state, the onset of superconductivity is higher than those in the paramagnetic (PM) state, but the transition is very broad.
This behavior is consistent with results by Wu {\it et al.} [1].
We defined the superconducting transition temperature $T_c$ for zero resistance.
At 0.52 (0.48) GPa, $T_c$ was estimated to be $0.8\pm0.4$ ($0.4\pm0.4$) from extrapolation, respectively.

\begin{figure}[htb]
\centering
\includegraphics[width=0.7\linewidth]{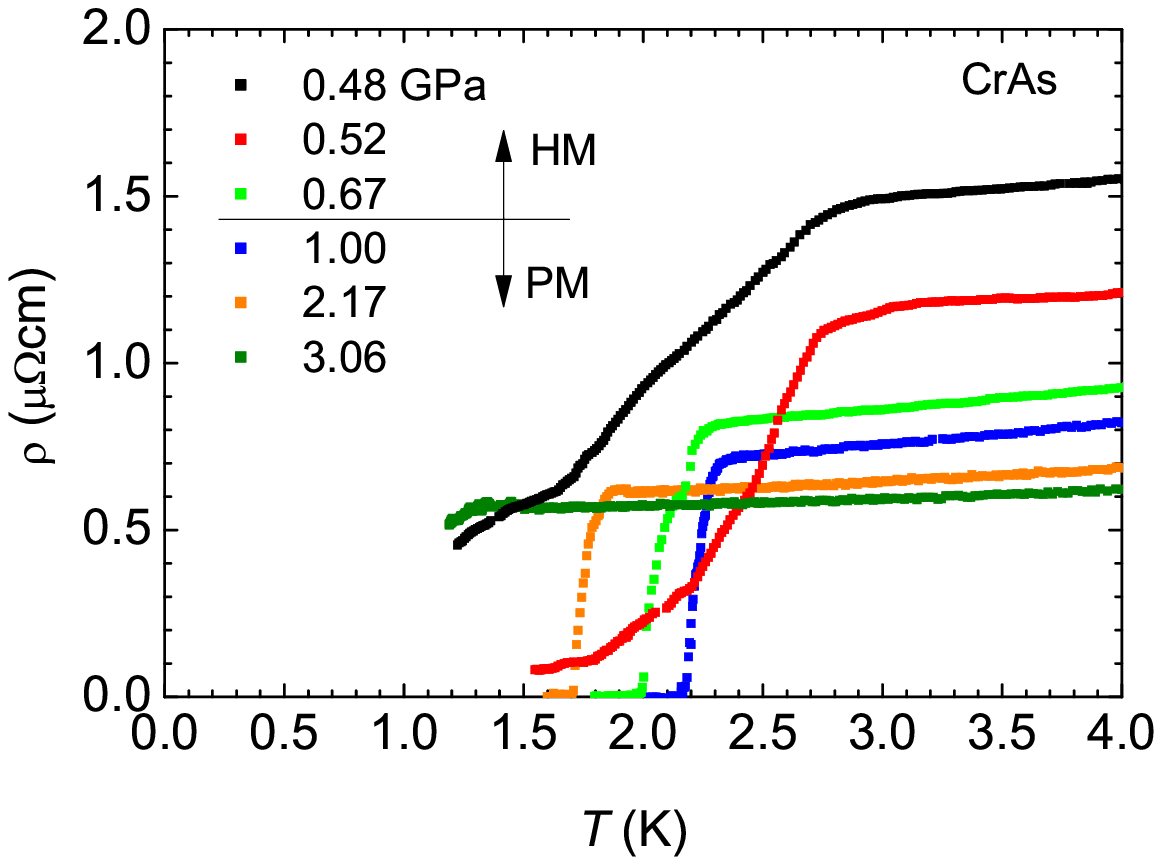}
\caption[]{(color online) Resistivity below 4 K measured at several pressures for CrAs .
}
\end{figure}

\vspace{15ex}

\subsection{Band calculation and Fermi surface}

Figs.~2(a-c) shows results of band calculation performed using the structural parameter at ambient pressure and room temperature [6].
The calculations were obtained through a full-potential LAPW (linear augmented plane wave) calculation within the LDA (local density approximation).
The density of states (DOS) shown in Fig.~2(b), which is almost consistent with a previous report [25].
In this figure, the red, green and blue curves indicate the partial DOS from Cr-$d$, As-$p$, and Cr-$p$ orbitals, respectively.
It indicates that the Fermi surface of CrAs is formed mainly by the Cr-$d$ orbitals and the peak DOS is located just above the Fermi energy, $E_F$.
As shown in Fig.~2(c), the electron Fermi surfaces have 3D character, whereas the hole surfaces are 2D-like and are almost degenerated.

\begin{figure}[htb]
\centering
\includegraphics[width=0.7\linewidth]{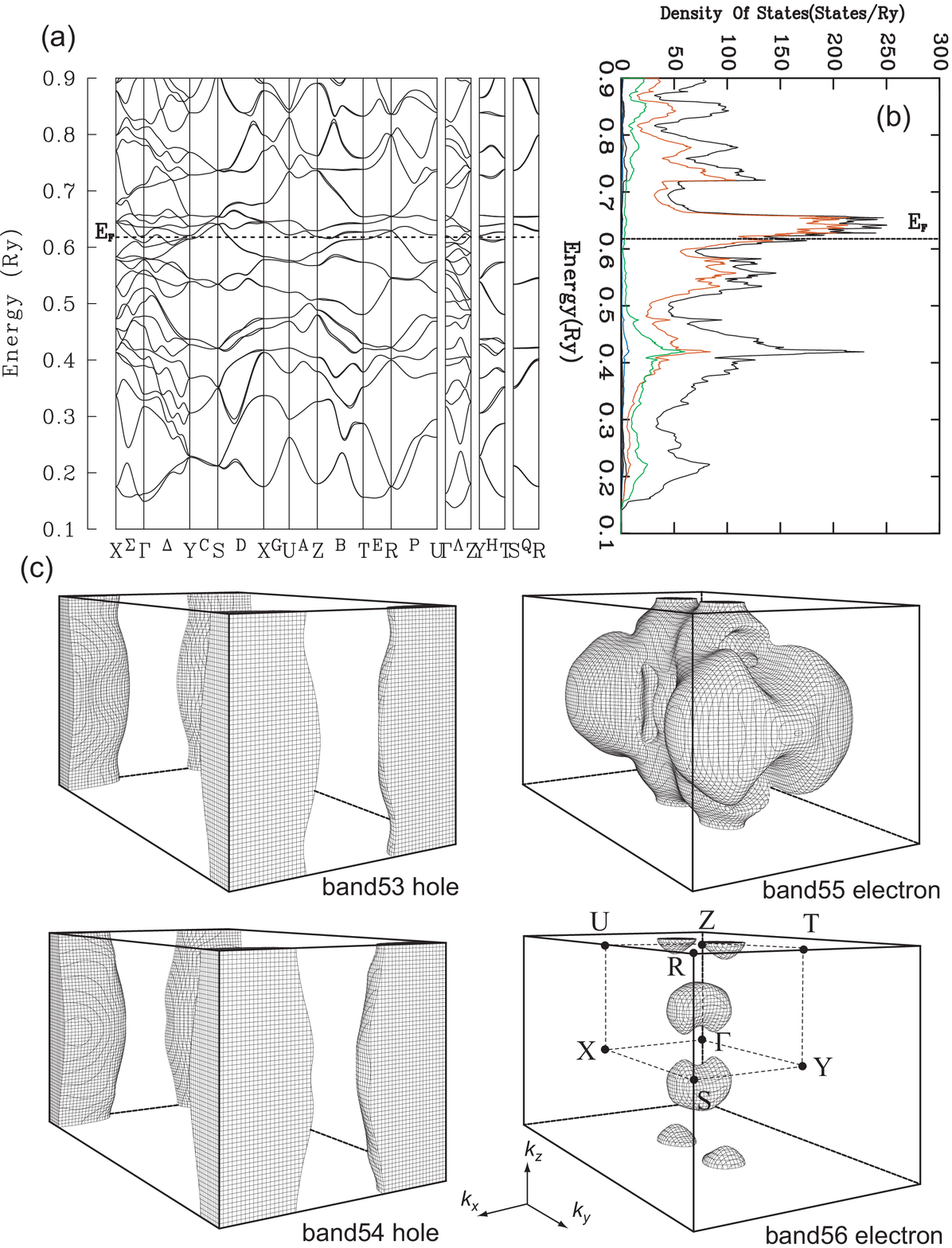}
\caption[]{(color online) (a) Band dispersion, (b) DOS, and (c) Fermi surfaces of CrAs calculated in the PM state.  
}
\end{figure}


\begin{references}
\bibitem{theta}
The $1/T_1T$ in the range of $20-280$ K is almost fitted with $-15 \leq \theta \leq 5$ for $\alpha=1$ and $-3 \leq \theta \leq 12$ for $\alpha=0.5$, but the fitting parameters are not obtained to reproduce the data down to low temperatures. We show a typical case of $\theta=0$ excepted at a QCP in Fig.~3.
\end{references}
\end{document}